\documentstyle[nature,doublespace]{article}
\begin{document}
\newcommand{\comment}[1]{}
\newcommand{\DGLOB}{D_{\rm glob}}
\newcommand{\DLOC}{D_{\rm loc}}
\newcommand{\DRANDGLOB}{D_{\rm glob}^{\rm random}}
\newcommand{\DRANDLOC}{D_{\rm loc}^{\rm random}}

\title{\bf Harmony in the Small-World}

\author{
Massimo Marchiori$^{1,2}$ and Vito Latora$^{3}$\\
\\ $^1$ The World Wide Web Consortium (W3C)\\ $^2$ MIT Lab for
Computer Science\\ 
$^3$ Dipartimento di Fisica Universit\'a di Catania \\ 
and INFN Sezione Catania}

\date{May 9, 2000} 
\begin{singlespace}
\maketitle
\end{singlespace}


\begin{abstract}
The Small-World phenomenon, popularly known
as six degrees of separation, has been mathematically
formalized by Watts and Strogatz in a study of the topological
properties of a network. Small-worlds
networks are defined in terms of two quantities: they
have a high clustering coefficient $C$ like regular lattices
and a short characteristic path length $L$ typical of random networks.
Physical distances are of fundamental importance 
in the applications to real cases, nevertheless this basic ingredient 
is missing in the original formulation. 
Here we introduce a new concept, the 
{\em connectivity~length~D\/}, that 
gives harmony to the whole theory. 
$D$ can be evaluated on a global and on a local scale 
and plays in turn the role of $L$ and $1/C$. 
Moreover it can be computed for any metrical network and 
not only for the topological cases.
$D$ has a precise meaning in term of information propagation and 
describes in an unified way both the structural and 
the dynamical aspects of a network: small-worlds are defined by a 
small global and local $D$, i.e. by a high efficiency in
propagating information both on a local and on a global scale. 

The neural system of the nematode {\em C.~elegans\/},  
the collaboration graph of  film actors, 
and the oldest U.S.\ subway system, can now 
be studied also as metrical networks and are shown to 
be small-worlds.

\end{abstract}

\bigskip
\newpage

Many biological and social systems present in nature, such as 
neural networks or nervous systems of living organisms, 
structured societies, but also internet and 
the world wide web, can be represented by means of a graph. 
In such a graph the vertices are the single units (the elements) 
of the system 
(neurons in the brain, human beings in a society, single computers 
in a computer network, etc.), while
the edges are the links or a measure of the interactions 
between the single elements. 
If we are interested in understanding the complex dynamical 
behaviour of many natural systems it is therefore very important 
to study the topological and the metrical properties 
of the underlying networks. 

The Small-World behaviour\cite{milgram,kochen}, popularly known
as six degrees of separation\cite{guare}, has been mathematically
formalized by Watts and Strogatz in a study of topological
networks\cite{watts,collins}.
The original formulation of small-world abstracts from
the real distances in a network: it only deals with the topology
of the connections, and imposes further constrains on the graph
structure, like total connectedness\cite{bollabas}.

In this paper we propose a general theory of small-world networks. 
We start by considering a generic metrical graph $G$, that 
in principle can be also non connected. 
$N$ is the number of vertices (or nodes) in the graph 
and $K$ is the total number of edges (or arcs). 
Each two nodes $i$ and $j$ of the graph are at a certain 
{\em physical distance\/} $\ell_{i,j}$, 
which can be for example the real distance between
the two nodes or a measure of the strength of their possible
interaction. The {\em distance on the graph\/}
$d_{i,j}$ is instead defined by the shortest
sum of the physical distances throughout all the possible
paths in the graph from $i$ to $j$. 
Let us suppose that every node sends information along 
the network, through its arcs. 
And, every node in the network propagates information concurrently. 
The amount of information sent from the node $i$ to the node $j$ per unity of
time is $v/d_{i,j}$, where $v$ is the velocity at which the
information travels over the network. When there is no path in
the network between $i$ and $j$, we assume
$d_{i,j}=+\infty$, and consistently, the amount of exchanged
information is $0$.
The {\em performance\/} of $G$ is the total amount of information
propagated per unity of time over the network: 
\begin{equation}
P=\sum_{i,j\in G} v/d_{i,j}
\end{equation}
Every sum here and in the following is intended for $i \neq j$. 
If we want to quantify the typical separation between two vertices in the
graph, a good measure is given by the 
{\em connectivity length\/} $D(G)$: this is the
fixed distance to which we have to set every two vertices in the
graph in order to maintain its performance.
Interestingly enough, the connectivity length of the graph is {\em
not the arithmetic mean\/}  but the {\em harmonic mean\/} of all
the distances:
\begin{equation}
\label{def}
D(G)=H(\{d_{i,j}\}_{i,j\in G})= \frac{N(N-1)} {\sum_{i,j\in G}
1/d_{i,j}}
\end{equation}

The harmonic mean has been known since the time of Pythagoras and
Plato\cite{timaeus} as the mean expressing ``harmonious and tuneful
ratios'', and later has been employed 
by musicians to formalize the diatonic scale, and 
by architects\cite{architecture} as paradigm for beautiful
proportions\cite{palladio}.
Nowadays, it finds extensive
applications in a variety of different fields, like 
traffic\cite{traffic},
information retrieval\cite{inforetrieval},
visibility systems\cite{asos},
water control\cite{water} and many others.
In particular, the harmonic mean is used to calculate
the average performance of computer systems\cite{performance1,benchmarkart},
parallel processors\cite{performance4}, and
communication devices (for example modems and
Ethernets\cite{performance2}). In all such cases, where a mean flow-rate
of information has to be computed, the simple arithmetic mean
gives the wrong result.

The definition of small-world proposed by Watts and Strogatz
is based on two different quantities:
a measure of the global properties of the graph ($L$),
defined as the average number of edges in the shortest path
between two vertices, and a local quantity ($C$),
measuring the average cliquishness properties of
a generic vertex. $C$ is the average number of
edges existing in the clique of a generic vertex (the graph of its
neighbors) divided by the maximum possible number of edges in the
clique. The main reason to introduce 
$C$ is because $L$, defined as the
simple arithmetic mean of $d_{i,j}$, applies only to connected
graphs and can not be used for cliques subgraphs, that in most of
the cases are disconnected.
In our theory we have a uniform description both of the 
global and of the local properties of the network 
by means of the single measure $D$. 
In fact we can evaluate $\DGLOB$, the connectivity length 
for the global graph $G$,
and $\DLOC$ the average connectivity length of its cliques.
Small-world networks are defined by small $D$ both on global and
local scale.
The connectivity length $D$ gives harmony to the whole theory of
small-world networks, since:

\begin{itemize}
\item
it is not just a generic intuitive notion of average distance in
a network, but has a precise meaning in terms of network
{\em efficiency\/}.

\item
it describes in an unified way the system on a global and on a
local scale.

\item
it applies both to topological and $metrical$ networks. ~The
topological approximation\cite{watts} is a too strong abstraction
for any real network. 
Here we show that small-worlds are indeed
present in nature, and are not just a topological effect.

\item
it applies to any graph, not only to connected graphs as
the original theory\cite{watts}.

%
\item
it describes the structural, but also the
$dynamical$ aspects of a network.
In the second part of their article, Watts and Strogatz
try to investigate the dynamics of a small-world network
by means of an example model of disease spreading\cite{watts}.
Using numerical simulations, they find out that the time of
disease propagation and $L$ have a similar form (see their
Fig.3b~).
In our theory this result and the very same concept of
effectiveness in the dynamics of signal propagation
in a small-world network are already implicit in the definition of $D$:
small $\DGLOB$ and $\DLOC$ mean a high performance both on
a local and on a global scale.
\end{itemize}

We now present a few numerical examples and some applications to 
real networks. 
As a first numerical experiment, we consider the original
topological example used by Watts and Strogatz \cite{watts}.
In fig.1 a small-world graph is constructed from a 
regular lattice with $N=1000$ vertices and $10$ edges per
vertex, by means a random rewiring process that 
introduces increasing disorder with probability $p$.
Such a procedure does not consider the geometry of
the system: in fact the physical distance $\ell_{i,j}$ between any two
vertices is always equal to 1.  
Here and in the following of the paper we calculate 
the distances on the graph $\{d_{i,j}\}_{i,j\in G}$ 
by means of the Floyd-Warshall algorithm\cite{pallottino}. 
This method is extremely efficient and allows to compute in parallel 
all the distances in the graph. Moreover $D$ is normalized to its minimum 
value $H(\{\ell_{i,j}\}_{i,j\in G})$ and ranges in $[1,+\infty]$. 
A similar normalization is also used in ref. \cite{watts}.
In panels a) and b) we plot the global and the local connectivity 
length versus p. 
The rewiring process produces small-world networks which result 
from the immediate drop in $\DGLOB$ caused by 
the introduction of a few long--range edges. 
Such ``short cuts'' connect vertices
that would otherwise be much farther apart at no cost
because $\ell_{i,j}=1 ~\forall i \neq j$.
During the drop of $\DGLOB$, $\DLOC$ remains
small and almost equal to the value for the regular lattice. 
Small-worlds are characterized by having small $\DGLOB$ and $\DLOC$,  
i.e. by high global and local performances.

We can now compare our definition with the original one 
given by Watts and Strogatz. 
In fig.2 we report the same quantities of fig.1 normalized by 
$\DGLOB(0)$ and $\DLOC(0)$, i.e. the values for the regular lattice.  
Our figure reproduces perfectly the results
shown in fig.2 of ref.\cite{watts}. With this normalization, 
the global and local connectivity lenghts behave respectively 
like $L$ and $1/C$, and in the topological 
case our theory gives the same results of Watts and Strogatz. 

The essential role of short cuts is emphasized by a second
experiment in which we repeat the test by changing the metric
of the system. 
In fig.3 we implement a random rewiring in which the length 
of each rewired edge is set to change from 1 to 3.  
This time short cuts have a cost. The figure shows that the 
small-world behaviour is still present even when the length of 
the rewired edges is larger than the original one and 
the behaviour of $D_{glob}$ is not simply monotonic decreasing. 
To check the robusteness of these results we increased
even more the length of the rewired edges, and we have 
tested many different metrical network (points on a circle
or on a bidimensional square lattice), obtaining similar
results.

This second numeric example suggests that the small-world behavior 
is not only an effect of the topological abstraction but can be 
found in nature in all such cases where the physical 
distance is important and the rewiring has a cost.
Therefore we have studied three real networks of great interest: 

\begin{itemize}
\item
the neural network of the nematode worm {\em C.~elegans\/} \cite{verme}, 
the only case of completely mapped neural network existing on the market. 

\item
the collaboration graph of actors in feature films\cite{actors}, which is 
an example of a non-connected social network.  

\item
the Massachusetts Bay underground transportation system\cite{tboston}, 
a case in which $\{\ell_{i,j}\}$ are given by the
space distances between stations $i$ and $j$. 
\end{itemize}

\noindent
These cases are interesting because they are all better described
by metrical rather than topological graphs. 

In table 1 we report the results on the {\em C.~elegans\/}.
 The {\em C.~elegans\/} nervous system consists of
$N=282$ neurons and two different types of links,
synaptic connections and gap junctions,
through which information is propagated from neuron to neuron.
In the topological case studied by Watts and Strogatz, the graph
consists of $K=2462$ edges, each defined when two vertices are
connected by either a synapse or a gap junction\cite{watts}. This
is only a crude approximation of the real network. Neurons are
different one from an other and some of them are in much stricter
relation than others: the number of junctions between two neurons
can vary a lot (up to a maximum of 72). A metrical network is more
suited to describe such a system and can be defined by setting
$\ell_{i,j}$ equal to the inverse number of junctions between $i$
and $j$. Connectivity lengths of real networks, compared to random 
graphs, show that the {\em C.~elegans\/} is both a topological and
a metrical small-world.

In table 2 we study the collaboration graph of actors 
extracted from the Internet Movie Database\cite{actors}, 
as of July 1999. The graph 
has $N=277336$ and $K=8721428$, and is not a connected graph.
The approach of Watts and Strogatz 
cannot be applied directly and they
have to restrict their analysis 
to the giant connected component of the graph\cite{watts}. 
Here we apply our small-world
analysis directly to the whole graph, without any restriction.
Moreover the topological case only provides whether actors 
participated in some movie together, or if they did not at
all. Of course, in reality there are instead various degrees of
correlation: two actors that have done ten movies together are 
in a much stricter relation rather than two actors that
have acted together only once. As in the case of {\em C.~elegans\/} we
can better shape this different degree of friendship by using a
metrical network: we set the distance $\ell_{i,j}$ between two
actors $i$ and $j$ as the inverse of the number of movies they did
together. The numerical values in table 2 indicate that 
both the topological and the metrical network 
show the small-world phenomenon.

The Massachusetts Bay transportation system\cite{tboston} ({\em MBTA\/},
popularly known as {\em T}) is the oldest subway system in the U.S.\
(the first electric streetcar line in Boston, which is now part of
the MBTA Green Line, began operation on January 1, 1889)
and consists of $N=124$ stations and $K=124$ tunnels 
extending through Boston and the other cities of the
Massachusetts Bay.
This is an example of an important real network where
the matrix $\ell_{i,j}$ is given by
the spatial distances between two stations $i$ and $j$ \cite{secondo}.
We have calculated such distances using information databases from
Geographic Data Technology (GDT), the U.S.\ Defence Mapping Agency,
and the National Mapping Division.
The comparison with random graphs in table 3 indicates
the MBTA is a small-world network, thus it is a very efficient
transportation system.
$D$ gives also precise quantitative information:
$D_{glob}=1.58$ shows that the MBTA is only 58\% less efficient
than the optimal subway ($D=1$) with a direct connection tunnel
for each couple of stations.
On the other side the relatively high value
of $D_{loc}$ indicates that the system is not perfectly
{\em fault tolerant\/}\cite{fault}.
This is intuitively explained by the fact
that usually most of the network is blocked if
a tunnel in the subway is interrupted.

To conclude, the values of $\DGLOB$ and $\DLOC$ in Tables 1-3 
show that all the three above cases (topological and metrical) 
are small-worlds.
These real examples, coming from different fields 
indicate that the small-world phenomenon
is not merely an artifact of an oversimplified topological model but a
common characteristic of biological and social systems.
The theory presented here provides with new general tool of
investigation for any complex system in nature
and it can be applied to a huge number of real cases.
Furthemore, it incorporates in a unified way
both the structural and the
dynamical aspects of a network.

\bigskip

{\bf Acknowledgments}
We thank Michel Baranger and Tim Berners-Lee
for many stimulating discussions and a
critical reading of the manuscript, Brett Tjaden for providing us with
the Internet Movie Database, and the MIT
Department of Biology for fruitful informations on {\em C.~elegans\/}.
One of us, V.L. thanks the MIT Center for Theoretical Physics  
and the Department of Physics of Harvard University, 
where the main part of this work has been done. 



\newpage
\pagestyle{empty}
\centerline{{\bf FIGURE CAPTIONS}}

\bigskip
Fig.1~~~ Topological case. We consider $N=1000$, $10$ edges per
vertex and the same random rewiring used by Watts and Strogatz
\cite{watts}. In panels a) and b) we plot the
global and the local connectivity length versus p.
The data shown in the figure are averages
over 10 random realizations of the rewiring process. 
The logarithmic horizontal scale is used to resolve the rapid drop in
$\DGLOB$ due to the presence of short cuts and corresponding to
the onset of the small-world.

\bigskip
Fig.2~~~ We normalize $\DGLOB$ and $\DLOC$ respectively 
by $\DGLOB(0)$ and $\DLOC(0)$ and we reproduce fig.2 of Watts and 
Strogatz \cite{watts}. $L$ is equivalent to 
$\DGLOB$, while $C$ plays the same role as $1/\DLOC$.

\bigskip
Fig.3~~~ Metrical case. We consider $N=1000$, $10$ edges per
vertex and we implement a random rewiring in which the length 
of each rewired edge is set to change from 1 to 3.
The data shown are averages over 10 random realizations 
of the rewiring process.


\newpage
\begin{tabular}{l|ll|ll}
\multicolumn{5}{l}{{\bf Table 1~~ C.\ elegans}}\\ \hline \\
 & $D_{glob}$ & $\DRANDGLOB$ & $D_{loc}$ & $\DRANDLOC$ \\ 
\hline
Topological  & 2.19 & 2.15 & 2.12 & 12.69\\ 
Metrical & 31.51 &31.28 & 8.64 & 36.93\\
\hline
\end{tabular}

\bigskip
\noindent
{\small 
Connectivity lengths of the real network  
compared to random graphs 
show that the {\em C.~elegans\/} is both a topological and
a metrical small-world.
}

\newpage
\begin{tabular}{l|ll|ll}
\multicolumn{5}{l}{{\bf Table 2~~ Film actors}}\\ \hline \\
 & $D_{glob}$ & $\DRANDGLOB$ & $D_{loc}$ & $\DRANDLOC$ \\ 
\hline
Topological & 3.37 & 3.01 & 1.87 & 1800 \\
Metrical &47.2 & 58.6 & 37.6 & 2600 \\
\hline
\end{tabular}

\bigskip
\noindent 
{\small 
Same as in table 1 for the collaboration graph of actors. 
Both the topological and the metrical graph show the small-world
phenomenon.
}

\newpage
\begin{tabular}{l|ll|ll}
\multicolumn{5}{l}{{\bf Table 3~~ T Boston}}\\ \hline \\
 & $D_{glob}$ & $\DRANDGLOB$ & $D_{loc}$ & $\DRANDLOC$ \\ \hline
Metrical & 1.58 & 11.06 & 25.87 & 250 \\
\hline
\end{tabular}

\bigskip
\noindent
{\small 
Connectivity lengths of the real network, compared to random graphs 
show that the Massachusetts Bay transportation system
is a metrical small-world. 
}

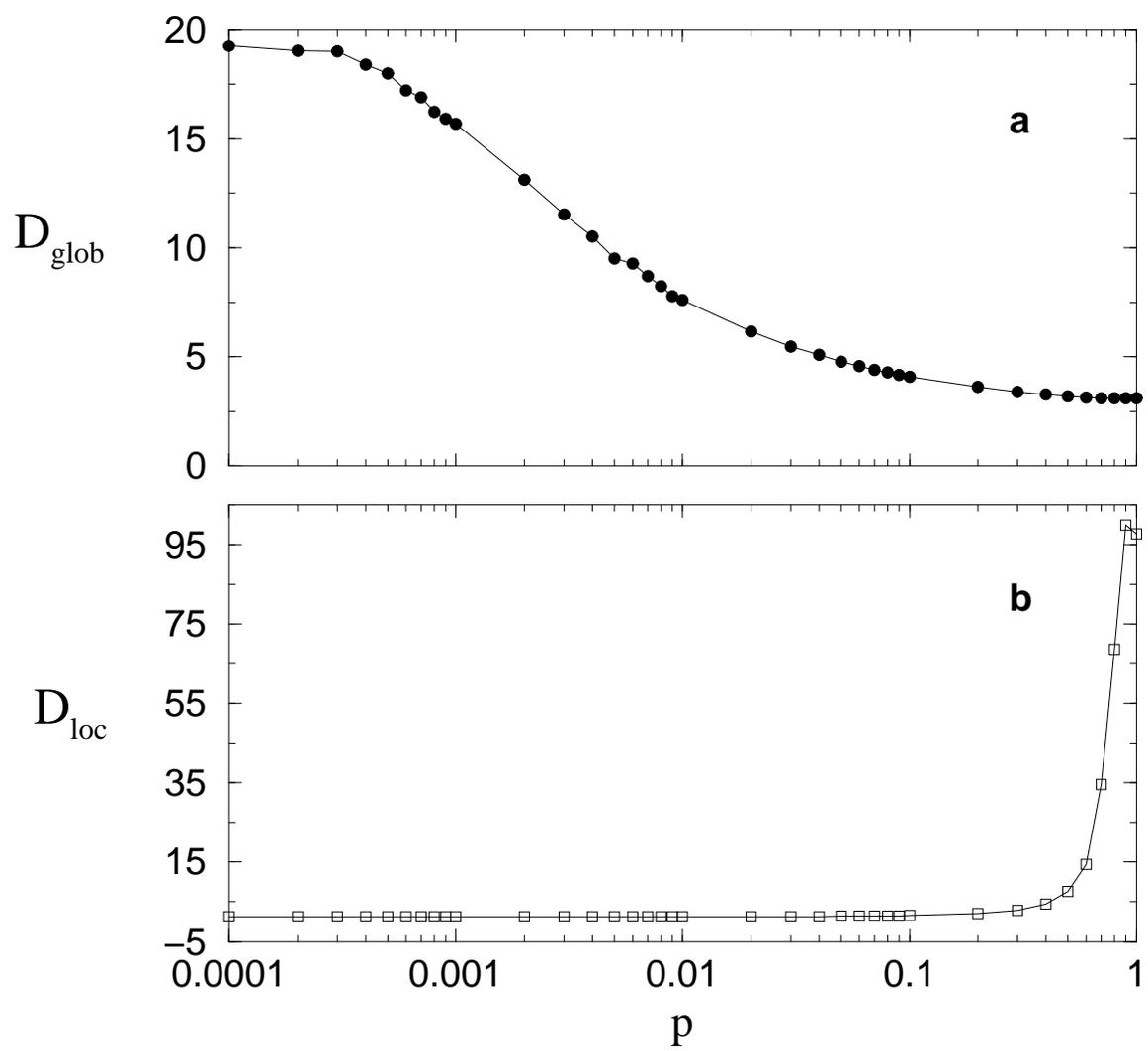
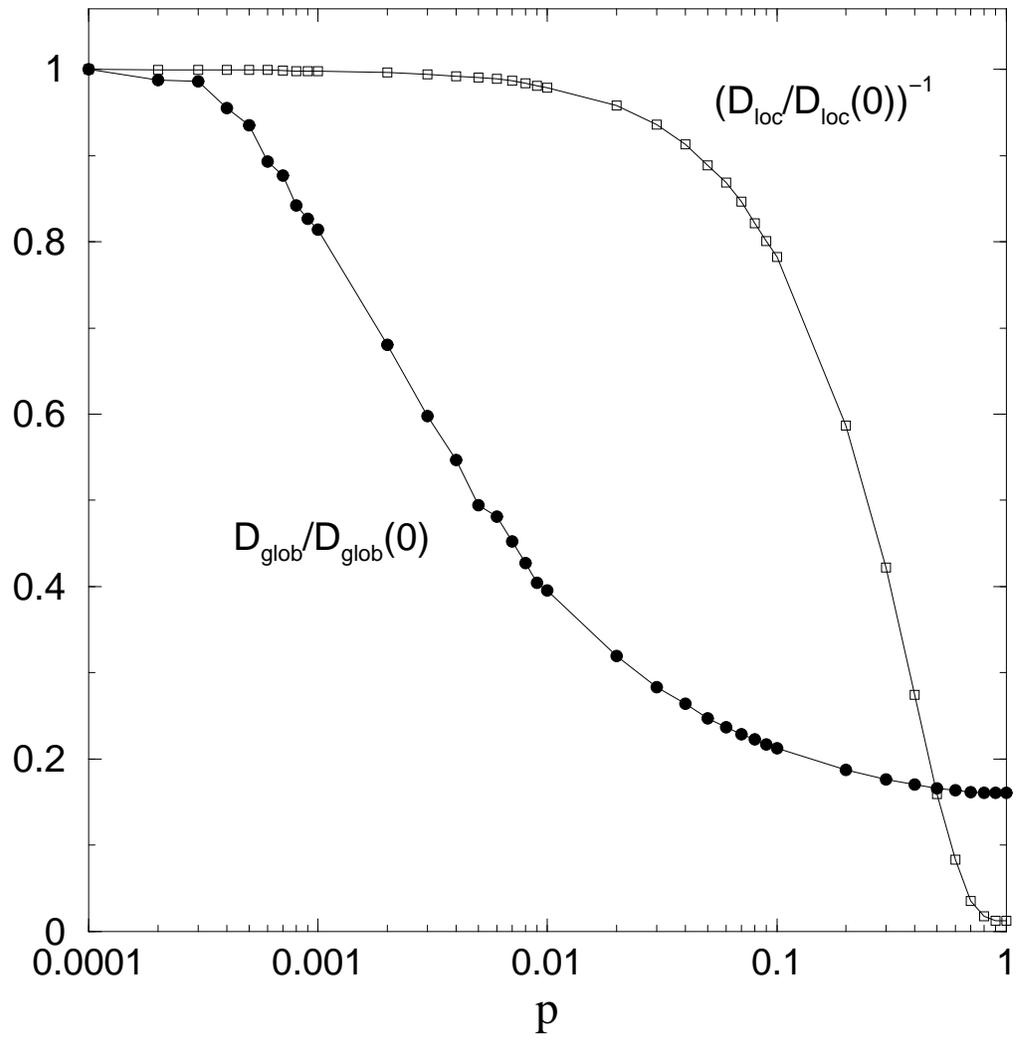
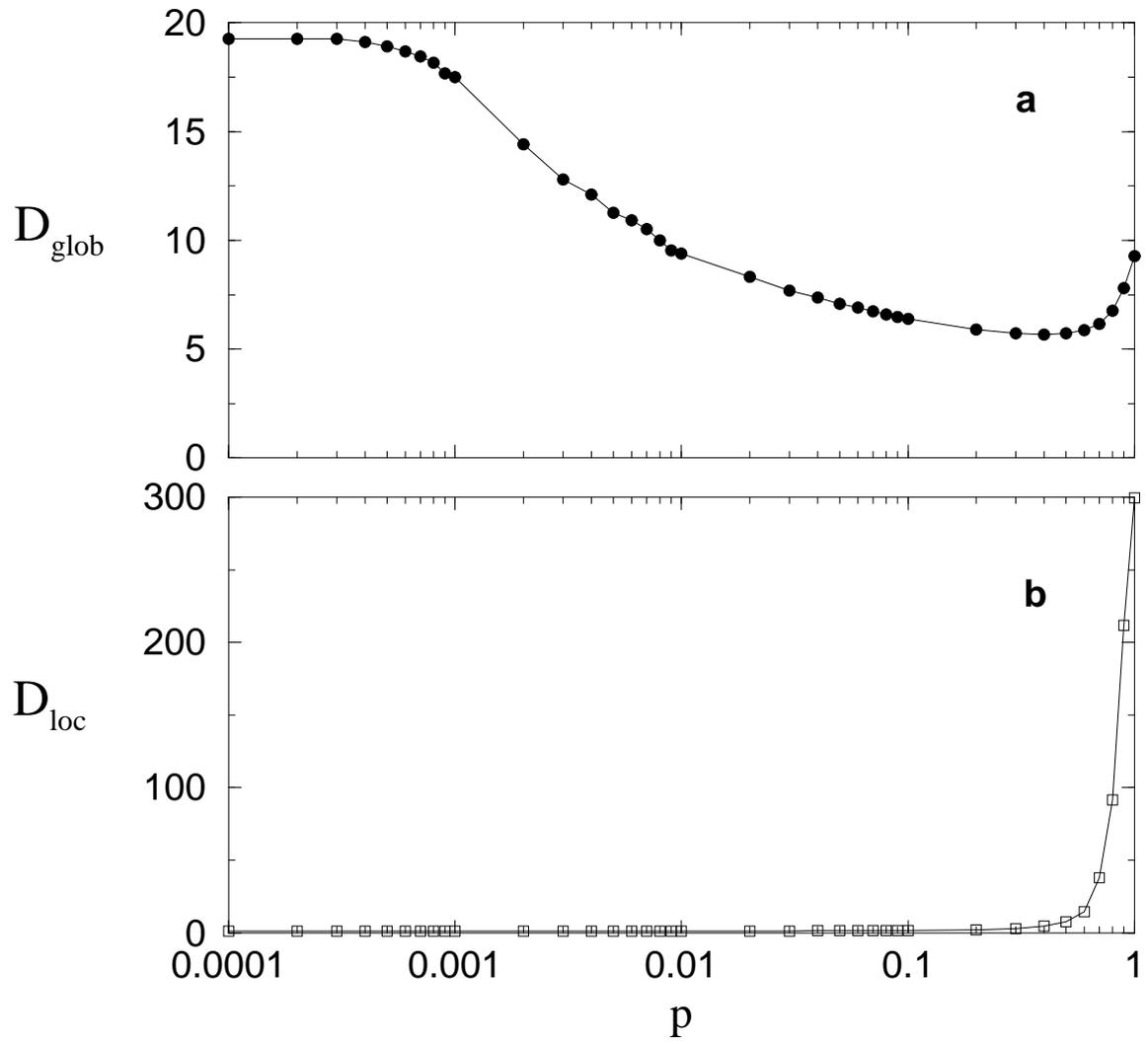
\end{document}